\documentclass[preprint,onecolumn,showpacs,preprintnumbers,amsmath,amssymb,times]{revtex4}

\usepackage{graphicx}
\usepackage{dcolumn}
\usepackage{bm}
\usepackage{psfrag}
\usepackage{epsfig}
\usepackage{amsmath}
\usepackage{amssymb}
\usepackage{color}
\usepackage{fancyheadings}
\usepackage{mathbbol}

\newcommand{\nn}{\nonumber}
\newcommand{\bra}{\langle}
\newcommand{\ve}{\vert}
\newcommand{\ket}{\rangle}
\newcommand{\iu}{\text{i}}
\newcommand{\ee}{\text{e}}
\newcommand{\lk}{\left}
\newcommand{\rk}{\right}

\begin{document}


\title{Excitation and Entanglement Transfer Near Quantum Critical Points}

\author{Michael J. Hartmann}
\email{m.hartmann@imperial.ac.uk}
\author{Moritz E. Reuter}
\author{Martin B. Plenio}
\affiliation{Institute for Mathematical Sciences, Imperial College London,
SW7 2PE, United Kingdom}
\affiliation{QOLS, The Blackett Laboratory, Imperial College London, Prince Consort Road,
SW7 2BW, United Kingdom}

\date{\today}

\begin{abstract}
Recently, there has been growing interest in employing condensed
matter systems such as quantum spin or harmonic chains as quantum
channels for short distance communication. Many properties of such
chains are determined by the spectral gap between their ground and
excited states. In particular this gap vanishes at critical points
of quantum phase transitions. In this article we study the
relation between the transfer speed and quality of such a system
and the size of its spectral gap. We find that the transfer is
almost perfect but slow for large spectral gaps and fast but
rather inefficient for small gaps.
\end{abstract}

\maketitle


\section{Introduction}

Quantum Information and Condensed Matter Physics are currently two
areas of very active research where several links between the two
branches have been discovered in recent years. In particular, the
relation of entanglement properties to critical points of quantum
phase transitions has been studied in some detail. In analogy to
classical phase transitions, the latter are typically
characterized in terms of the scaling behavior of equilibrium
properties, such as the correlation length. Analogous scaling
phenomena have now been found for the entanglement properties of a
spin chain in the vicinity of a critical point.
\cite{Osterloh2002,locent,JL05,ADA}.

In Quantum Communication, condensed matter systems have recently
received some attention as interesting candidates for quantum
channels \cite{Bos03,PHE04}. This approach may be of advantage in
situations where photonic quantum communication is not possible
for example because the two units are only separated by a few
optical wavelengths or are inside a material that does not permit
for the well-controlled propagation of light.

The quantum mechanical properties of correlated many body system
are to some extent determined by the energetic gap between their
ground states and the lowest excited states. In particular the
system is short range correlated if it features a finite gap,
while some quantities can become long range correlated when the
gap closes \cite{Has,EC05}. The latter happens at the critical
points of quantum phase transitions \cite{Sachdev1999}.

Motivated by these findings the dynamical entanglement properties
of quantum many body systems undergoing a quantum phase transition
are receiving increasing attention \cite{AUL05,YCW05,ZDZ05,HRP06}.

To pursue the analysis of quantum many body systems as possible
quantum channels a step further, we study here the connection
between the spectral gap and the systems ability to transfer
excitations and entanglement. However, our aim is twofold: Besides
being interested in the systems transfer capacity, we also want to
explore, to what degree the transfer properties can be employed to
detect and characterize critical points of quantum phase
transitions.

In particular, we consider spin chains and harmonic chains,
where in both cases, two ancillas couple weakly to the chain at distant sites.
We study the transfer of quantum information and excitations from one
ancilla to the other numerically, by employing newly developed matrix product state
techniques \cite{Vid03}, and analytically with a master equation approach.
The numerics is applied to the spin chains while the master equation is used
to describe the case of an harmonic chain, where its validity is confirmed
numerically for Gaussian initial states \cite{PHE04}.

Both systems show a very similar behavior: The transfer properties
sensitively depend on the energy gap between the ground state and
the lowest excited states, but do not significantly dependent on
the detailed structure of the Hamiltonian. The quality and speed
of state transfer through such systems may thus be used to detect
critical points experimentally.

This paper is organized as follows: Section II begins with a
discussion of spin chains, mainly employing numerical methods. In
Section III a simplified picture unraveling the origin of the
observed behavior is presented. This picture is then further
corroborated in Section IV where we study harmonic chains both
numerically and analytically with a master equation approach.
Section V discusses connection to quantum channel capacities and
Section VI summarizes our results.

%
\section{Spin chains}
A variety of quantum spin chains exhibit quantum phase
transitions. Therefore we begin with the study of a linear chain
of spins with nearest neighbor interactions and open boundary
conditions described by the Hamiltonian
\begin{equation}\label{hamchain}
H_{\textrm{chain}} = B \sum_{i=1}^N \sigma_i^z + \sum_{i=1}^{N-1}
\lk( J_x \sigma_i^x \sigma_{i+1}^x + J_y \sigma_i^y \sigma_{i+1}^y
+ J_z \sigma_i^z \sigma_{i+1}^z \rk).
\end{equation}
Here $N$ is the number of spins, $B$ a magnetic field and $J_x$,
$J_y$ and $J_z$ the interaction between neighboring spins. Two
ancilla systems (named $S$ for ``sender'' and $R$ for
``receiver'') couple to the chain at spins $m_S$ and $m_R$, which
are near the center of the chain in order to avoid boundary
effects. The total Hamiltonian of chain and ancillas then reads
\begin{equation}\label{hamtot}
H =
H_{\textrm{chain}} + B_a \lk( \sigma_S^z + \sigma_R^z \rk) + J_a
\lk( \sigma_S^x \sigma_{m_S}^x + \sigma_R^x \sigma_{m_R}^x \rk) \, ,
\end{equation}
where $B_a \ge 0$ is the Zeeman splitting of the ancillas, which might
differ from $B$, and $J_a \ge 0$ the coupling of the ancillas
to the chain, which is taken to be weak, i.e. $J_a \ll (B, J_x, J_y, J_z)$.
As an initial state of the system we consider
\begin{equation}
\ve \Psi (0) \ket = \ve \uparrow_S, \downarrow_R, 0_{\textrm{chain}} \ket \, ,
\end{equation}
i.e. the sender is spin up and the receiver is spin down, while
the chain is in its ground state, $\ve 0_{\textrm{chain}} \ket$.
This will allow us to explore the propagation of this excitation
through the chain.

The dynamics is simulated with a matrix product state technique
using matrices of dimension $10 \times 10$ \cite{Vid03,PED+}. The
accuracy of the simulations was confirmed by varying the matrix
dimension and the size of the time steps. Furthermore, we
confirmed that the dynamics conserves the total energy of the
system.

\begin{figure}
\psfrag{t}{\raisebox{-0.4cm}{$t$}}
\psfrag{1}{\hspace{-0.38cm}\small{$1.0$}}
\psfrag{0.1}{\hspace{-0.3cm}\small{$0.0$}}
\psfrag{0.2}{\hspace{-0.3cm}\small{$0.2$}}
\psfrag{0.4}{\hspace{-0.3cm}\small{$0.4$}}
\psfrag{0.6}{\hspace{-0.3cm}\small{$0.6$}}
\psfrag{0.8}{\hspace{-0.3cm}\small{$0.8$}}
\psfrag{0}{\raisebox{-0.2cm}{\small{$0$}}}
\psfrag{500}{\raisebox{-0.1cm}{\small{$ $}}}
\psfrag{1000}{\raisebox{-0.2cm}{\hspace{-0.2cm}\small{$5 \times 10^4$}}}
\psfrag{1500}{\raisebox{-0.1cm}{\small{$ $}}}
\psfrag{2000}{\raisebox{-0.2cm}{\hspace{-0.25cm}\small{$10 \times 10^4$}}}
\psfrag{2500}{\raisebox{-0.1cm}{\small{$ $}}}
\psfrag{3000}{\raisebox{-0.2cm}{\hspace{-0.25cm}\small{$15 \times 10^4$}}}
\includegraphics[width=7cm]{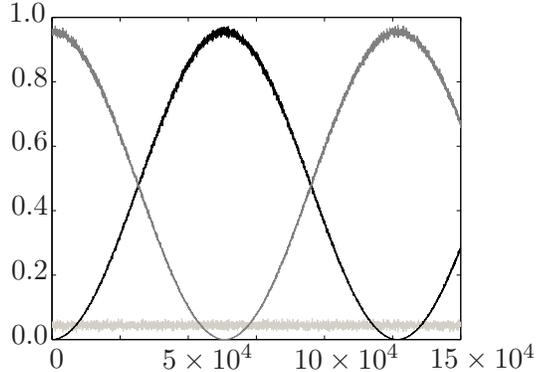}
\caption{\label{simulnonres} $P(\downarrow_S \downarrow_R)(t)$ (light gray), $P(\uparrow_S \downarrow_R)(t)$
(gray) and $P(\downarrow_S \uparrow_R)(t)$ (black) for $B = 1$, $J_x = 0.3$, $J_y = J_z = 0$, $B_a = 0.64$ and $J_a = 0.05$
as given by the simulation for the open boundary model with $N = 100$ spins.
$S$ couples to spin $45$ and $R$ to spin $55$.
Figure taken from \cite{HRP06}.}
\end{figure}

Figure \ref{simulnonres} shows the probability  $P(\uparrow_S
\downarrow_R)$ that ``sender'' $S$ is in its excited state $\ve
\uparrow_S \ket$ and the ``receiver'' $R$ in its ground state $\ve
\downarrow_R \ket$, together with $P(\downarrow_S \uparrow_R)$ and $P(\downarrow_S \downarrow_R)$ for a model with $N = 100$, $m_S = 45$, $m_R = 55$, $B =
1$, $J_x = 0.3$, $J_y = J_z = 0$, $B_a = 0.64$ and $J_a = 0.05$. $P(\uparrow_S \uparrow_R)$ is always less than
$10^{-4}$.
%
%
The plots show that the excitation that was initially located in
$S$ oscillates back and forth between $S$ and $R$.

\begin{figure}
\psfrag{t}{\raisebox{-0.4cm}{$t$}}
\psfrag{1}{\hspace{-0.38cm}\small{$1.0$}}
\psfrag{0.1}{\hspace{-0.3cm}\small{$0.0$}}
\psfrag{0.2}{\hspace{-0.3cm}\small{$0.2$}}
\psfrag{0.4}{\hspace{-0.3cm}\small{$0.4$}}
\psfrag{0.6}{\hspace{-0.3cm}\small{$0.6$}}
\psfrag{0.8}{\hspace{-0.3cm}\small{$0.8$}}
\psfrag{0}{\raisebox{-0.2cm}{\small{$0$}}}
\psfrag{200}{\raisebox{-0.2cm}{\hspace{0.0cm}\small{$200$}}}
\psfrag{400}{\raisebox{-0.2cm}{\hspace{0.0cm}\small{$400$}}}
\psfrag{600}{\raisebox{-0.2cm}{\hspace{0.0cm}\small{$600$}}}
\psfrag{800}{\raisebox{-0.2cm}{\hspace{0.0cm}\small{$800$}}}
\psfrag{1000}{\raisebox{-0.2cm}{\hspace{0.0cm}\small{$1000$}}}
\includegraphics[width=7cm]{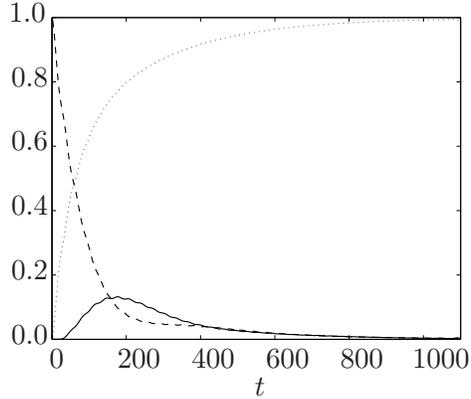}
\caption{\label{simulres} $P(\downarrow_S \downarrow_R)(t)$ (dotted line), $P(\uparrow_S \downarrow_R)(t)$ (dashed line)
and $P(\downarrow_S \uparrow_R)(t)$ (solid line) for $B = 1$, $J_x = 0.3$, $J_y = J_z = 0$, $B_a = 0.8$ and
$J_a = 0.05$ as given by the simulation for the open boundary model with $N = 600$ spins.
$S$ couples to spin $295$ and $R$ to spin $305$. Figure taken from \cite{HRP06}.}
\end{figure}

Figure \ref{simulres} shows $P(\uparrow_S \downarrow_R)$, $P(\downarrow_S \uparrow_R)$
and $P(\downarrow_S \downarrow_R)$ for
a model with $N = 600$, $m_S = 295$, $m_R = 305$, $B = 1$, $J_x = 0.3$, $J_y = J_z = 0$, $B_a = 0.8$ and $J_a = 0.05$.
Again, $P(\uparrow_S \uparrow_R)$ is always less than
$10^{-4}$.
%
%
For these parameters, the excitation is not fully transferred to
$R$, contrary to figure \ref{simulnonres}. Both, $S$ and $R$ relax
to their ground states with the excitation only being partially
and temporarily transferred to $R$, even for close-lying spins.
Note that the parameters chosen in figures \ref{simulnonres} and \ref{simulres}
are the same except for $B_a$ which in figure \ref{simulres} is significantly larger
than in figure \ref{simulnonres}.

\begin{figure}
\psfrag{t}{\raisebox{-0.4cm}{$t$}}
\psfrag{1}{\hspace{-0.44cm}\small{$1.0$}}
\psfrag{0.1}{\hspace{-0.4cm}\small{$0.0$}}
\psfrag{0.2}{\hspace{-0.3cm}\small{$0.2$}}
\psfrag{0.4}{\hspace{-0.3cm}\small{$0.4$}}
\psfrag{0.6}{\hspace{-0.3cm}\small{$0.6$}}
\psfrag{0.8}{\hspace{-0.3cm}\small{$0.8$}}
\psfrag{0}{\raisebox{-0.2cm}{\small{$0$}}}
\psfrag{2000}{\raisebox{-0.2cm}{\hspace{-0.0cm}\small{$2000$}}}
\psfrag{4000}{\raisebox{-0.2cm}{\hspace{-0.0cm}\small{$4000$}}}
\psfrag{6000}{\raisebox{-0.2cm}{\hspace{-0.0cm}\small{$6000$}}}
\psfrag{8000}{\raisebox{-0.2cm}{\hspace{-0.0cm}\small{$8000$}}}
\psfrag{10000}{\raisebox{-0.2cm}{\hspace{-0.0cm}\small{$10000$}}}
\includegraphics[width=7cm]{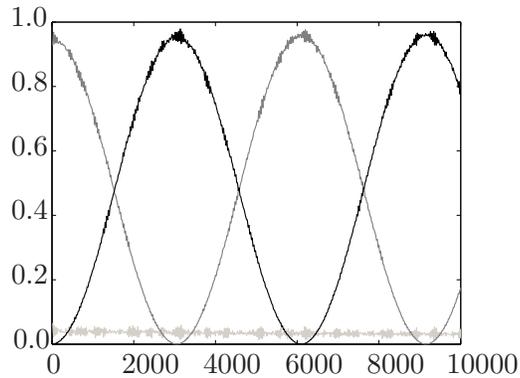}
\caption{\label{simulnonres1} $P(\downarrow_S \downarrow_R)(t)$ (light gray),
$P(\uparrow_S \downarrow_R)(t)$ (gray)
and $P(\downarrow_S \uparrow_R)(t)$ (black)
for $B = 1$, $J_x = 0.5$, $J_y = 0.2$, $J_z = 0.1$, $B_a = 0.04$ and $J_a = 0.05$
as given by the simulation for the open boundary model with $N = 100$ spins.
$S$ couples to spin $45$ and $R$ to spin $55$. Figure taken from \cite{HRP06}.}
\end{figure}

\begin{figure}
\psfrag{t}{\raisebox{-0.4cm}{$t$}}
\psfrag{1}{\hspace{-0.38cm}\small{$1.0$}}
\psfrag{0.1}{\hspace{-0.3cm}\small{$0.0$}}
\psfrag{0.2}{\hspace{-0.3cm}\small{$0.2$}}
\psfrag{0.4}{\hspace{-0.3cm}\small{$0.4$}}
\psfrag{0.6}{\hspace{-0.3cm}\small{$0.6$}}
\psfrag{0.8}{\hspace{-0.3cm}\small{$0.8$}}
\psfrag{0}{\raisebox{-0.2cm}{\small{$0$}}}
\psfrag{100}{\raisebox{-0.2cm}{\small{$200$}}}
\psfrag{200}{\raisebox{-0.2cm}{\small{$400$}}}
\psfrag{300}{\raisebox{-0.2cm}{\small{$600$}}}
\includegraphics[width=7cm]{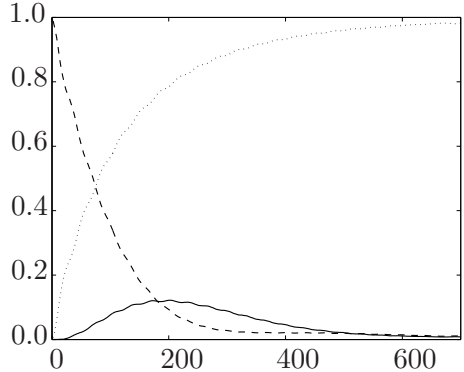}
\caption{\label{simulres1} $P(\downarrow_S \downarrow_R)(t)$ (dotted line),
$P(\uparrow_S \downarrow_R)(t)$ (dashed line)
and $P(\downarrow_S \uparrow_R)(t)$ (solid line)
for $B = 1$, $J_x = 0.3$, $J_y = 0.2$, $J_z = 0.1$, $B_a = 0.2$ and $J_a = 0.05$
as given by the simulation for the open boundary model with $N = 600$ spins.
$S$ couples to spin $295$ and $R$ to spin $305$. Figure taken from \cite{HRP06}.}
\end{figure}

The two observed scenarios are rather generic. To demonstrate this,
we have done the same simulations for a XXZ-model.
The results, shown in figures \ref{simulnonres1} and \ref{simulres1},
clearly agree with our findings for the previous coupling parameters.
Again $B_a$ in figure \ref{simulres1} is significantly larger
than in figure \ref{simulnonres1}, while all other parameters are equal.

%
\section{Simplified physical picture}
\begin{figure}
\psfrag{S}{\raisebox{-0.2cm}{\hspace{-0.1cm}$S$}}
\psfrag{R}{\raisebox{-0.2cm}{\hspace{-0.1cm}$R$}}
\psfrag{Chain}{\raisebox{-0.2cm}{\hspace{-0.1cm}chain}}
\psfrag{Re0}{\raisebox{0.1cm}{\hspace{-0.6cm}$ $}}
\psfrag{Ren0}{\raisebox{0.1cm}{\hspace{-0.6cm}$ $}}
\includegraphics[width=3cm]{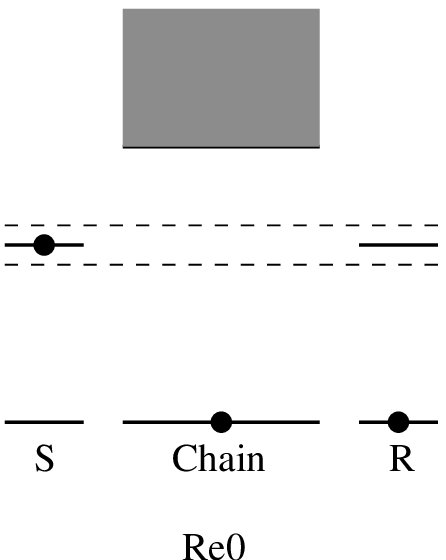}
\hspace{1cm}
\includegraphics[width=3cm]{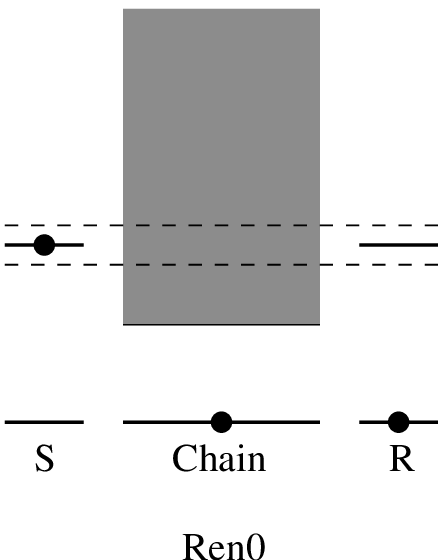}
\caption{\label{energycon} Sketch of the energy levels of the system. The dots indicate the occupations
of the initial state. For this initial state only the energy levels between the two horizontal
dashed lines are accessible, resulting in almost perfect transfer for the left scenario
and damping for the right one.}
\end{figure}

The exact dynamics of a quantum spin chain is extraordinarily
complex. Nevertheless, underlying the dramatic difference between
the almost perfect transfer scenarios in figures \ref{simulnonres}
and \ref{simulnonres1} and the damped scenario in figures
\ref{simulres} and \ref{simulres1} is a simple physical mechanism.
The dynamics we have simulated is given by the Schr\"odinger
equation containing the Hamiltonian (\ref{hamtot}). As a
consequence, all moments of the Hamiltonian are conserved.
The initial state $\ve \Psi (0) \ket$, where one of the ancillas is excited,
is not an eigenstate of $H$ as given by (\ref{hamtot}).
The mean value is $\bra \Psi (0) \ve H \ve \Psi (0) \ket$
and the variance of the energy is given by
\begin{equation}
\Delta E = \sqrt{\bra \Psi (0) \ve H^2 \ve \Psi (0) \ket - \bra
\Psi (0) \ve H \ve \Psi (0) \ket^2} = \sqrt{2} \, J_a \, .
\end{equation}
Given that the evolution is assumed to be Hamiltonian, both
are in fact time invariant.
Thus in the entire dynamics only those states with an energy
expectation value $\overline{E}$ in the range $\bra \Psi (0) \ve H
\ve \Psi (0) \ket - \sqrt{2} J_a < \overline{E} < \bra \Psi (0)
\ve H \ve \Psi (0) \ket + \sqrt{2} J_a$ play a significant role as
indicated by Figure \ref{energycon}. $S$ and $R$ are depicted as
two level systems, while for the chain there is a unique ground
state and a quasi continuous band of excited states sketched as
the gray area. The dots indicate the initial occupations. The
relevant energy range given by the variance lies between the two
horizontal dashed lines. If the spectral gap is larger than the
Zeeman splitting of the ancillas (left plot) it is not possible to
create a real excitation of the chain, only processes involving
virtual excitations of the chain are then permitted and the
excitation will be almost completely transferred from $S$ to $R$.
If however the spectral gap is smaller than the Zeeman splitting
of the ancillas, real excitations may be created in the chain.
These excitations dephase rapidly due to the large
number of accessible energy levels which prevents their
reabsorption -- the chain acts as a bath and information is lost.

We believe that this heuristic picture captures the essential
physics. To further corroborate our simple physical model and to
underline the generality of our findings, we now turn to a
different model for the chain which also features an adjustable
energy gap above its unique ground state.

%
\section{Harmonic chain}
Harmonic chains whose time evolution is governed by a Hamiltonian
that is quadratic in position and momentum can be solved exactly
in a compact form and therefore lend themselves to analytical
investigations. Furthermore, Gaussian initial states, ie states
whose Wigner function is a Gaussian, preserve their Gaussian
character under the time evolution. As a Gaussian is fully
described by its first and second moments this allows for a very
compact, polynomial in the number of oscillators, description of
the state of the systems and its evolution in time (see e.g.
\cite{Eisert P 03} for details), which permits the numerical
investigation of very large systems. This is the basic motivation
for the study of harmonic systems in Gaussian states in this
section.

We consider a harmonic chain with periodic boundary conditions
described by
\begin{equation} \label{harmchain}
H_{\textrm{chain}} = \frac{1}{2} \sum_{j=1}^N \lk( p_j^2 + \Omega^2 (q_j - q_{j+1})^2 +
\Omega_0^2 q_j^2 \rk)
\end{equation}
with the $p_j$ being the momenta and the $q_j$ the positions
($q_{N+1} = q_1$). In this case, the two ancillas are harmonic
oscillators that couple to oscillators $m_S$ and $m_R$ of the
chain. The complete Hamiltonian reads
\begin{eqnarray} \label{harmtot}
H_{\textrm{tot}} & = & H_{\textrm{chain}} + H_{\textrm{ancillas}} + H_I \\
H_{\textrm{ancillas}} & = & \frac{1}{2} \lk( p_S^2 + \omega^2 q_S^2 + p_R^2 + \omega^2 q_R^2 \rk) \\
H_I & = & J_a (q_S q_{m_S} + q_R q_{m_R}) \, .
\end{eqnarray}
We are interested in the time evolution of the ancillas only.
Thus, we will now derive an approximate master equation for the
reduced density matrix $\sigma(t)$ of the two ancilla systems. For
weak coupling $J_a \ll (\Omega, \Omega_0)$ we find
\begin{equation}\label{nakajima}
\frac{d\sigma}{dt} = - \int_0^t ds \, \textrm{Tr}_{\textrm{chain}}
\lk\{ \lk[ H_I(t), \lk[ H_I(s), \ve 0 \ket \bra 0 \ve \otimes
\sigma(s) \rk] \rk] \rk\} \, ,
\end{equation}
where $\sigma(t)$ is the reduced density matrix of the ancillas,
$\textrm{Tr}_{\textrm{chain}}$ is the trace over the degrees of
freedom of the chain and $\ve 0 \ket$ denotes the ground state of
the chain (\ref{harmchain}). $H_I(t)$ is the interaction between
ancillas and chain in the interaction picture. Here $H = H_0 +
H_I$ with $H_0 = H_{\textrm{chain}} + H_{\textrm{ancillas}}$ and
$$H_I(t) = e^{i H_0 t} H_I e^{- i H_0 t}$$ and $$\sigma(t) = e^{i
H_0 t} \rho e^{- i H_0 t}.$$
Eq. (\ref{nakajima}) is an expansion in the
coupling strength $J_a$ up to second order,
which is a good approximation if the integral approaches a
constant value for $t > t^{\star}$, where $J_a t^{\star} \ll 1$.
Since $\sigma$ only changes significantly on time scales
$t \sim J_a^{-1} \gg t^{\star}$, the approximation
$\sigma(s) \approx \sigma(t)$ can be used. Performing the
trace on the rhs of (\ref{nakajima}) yields
\begin{eqnarray} \label{masterform}
\frac{d\sigma}{dt} = - J_a^2 \sum_{j, l = S, R} &&\lk(
- i \lk(Y_1 + (Y_0 - Y_1)\delta_{jl}\rk) [a_j a_l^{\dagger}, \sigma] \rk. \nn \\
& & \enspace + \lk. \lk(X_1 + (X_0 - X_1)\delta_{jl}\rk) \lk( \{ a_j a_l^{\dagger}, \sigma \}
- 2 ( a_j \sigma a_l^{\dagger} )\rk) \rk) \, ,
\end{eqnarray}
where $[ \cdot, \cdot ]$ and $\{ \cdot, \cdot \}$ denote
commutators and anti-commutators and $a_S$ and $a_R$ are the
annihilation operators of $S$ and $R$, respectively
\begin{eqnarray*}
q_j & = & \frac{1}{\sqrt{2 \omega}}(a_j + a_j^{\dagger})\\
p_j & = & -\iu \sqrt{\frac{\omega}{2}} (a_j - a_j^{\dagger})
\end{eqnarray*}
for $j = S, R$. In deriving (\ref{masterform}), we applied a
rotating wave approximation. The coefficients read
\begin{eqnarray*}
    X_0 &=& \frac{1}{2\omega}\textrm{Re}(C_{m_S m_S}^+ + C_{m_S m_S}^-),\\
    X_1 &=& \frac{1}{2\omega}\textrm{Re}(C_{m_S m_R}^+ + C_{m_S m_R}^-),\\
    Y_0 &=& \frac{1}{2\omega}\textrm{Im}(C_{m_S m_S}^+ + C_{m_S m_S}^-),\\
    Y_1 &=& \frac{1}{2\omega}\textrm{Im}(C_{m_S m_R}^+ + C_{m_S m_R}^-)
\end{eqnarray*}
with $C_{kl}^{\pm}$ given by
\begin{equation} \label{mastercoeff}
C_{kl}^{\pm} (t) = \int_0^t ds \, \bra 0 \ve q_k (t) q_l (s) \ve 0 \ket \,
\ee^{\pm i \omega (t-s)} \, ,
\end{equation}
where $k, l = m_S, m_R$.
Due to the symmetries of the model, we have $C_{m_S m_S}^{\pm} = C_{m_R m_R}^{\pm}$ and
$C_{m_S m_R}^{\pm} = C_{m_R m_S}^{\pm}$.
Eq. (\ref{masterform}) is a good approximation whenever
$C_{kl}^{\pm} (t) \approx \overline{C}_{kl}^{\pm} =
\textrm{const.} \quad \textrm{for} \quad t \ll J_a^{-1}$.
Since the $C_{kl}^{\pm} (t)$ do not depend on $J_a$ themselves,
there is always a sufficiently small $J_a$ such that this holds,
provided $\lim_{t \to \infty} C_{kl}^{\pm} (t)$ exists. The
validity of the master equation can be confirmed by a numerical
simulation for a chain with 1400 oscillators. We found good
agreement between our analytical and numerical solutions, with the
relative errors being less than $5\%$.

The harmonic chain can be diagonalized via a Fourier transform \cite{PHE04}.
In the limit of an infinitely long chain, $N \rightarrow \infty$,
the correlation functions read
\begin{equation} \label{correlfun}
\bra 0 \ve q_j (t) q_l (s) \ve 0 \ket = \frac{1}{2 \pi}
\int_0^{\pi} dk \, \frac{1}{\omega_k} \, \cos((j - l) k) \, \ee^{-
i \omega_k (t - s)} \, ,
\end{equation}
where $\omega_k^2 = 4 \Omega^2 \sin^2 (k / 2) + \Omega_0^2$ with
$- \pi < k < \pi$. and all $\lim_{t \to \infty} C_{kl}^{\pm} (t)$
exist except for the case where $\omega = \Omega_0 = 0$. As in
master equations for system bath models, we now insert the
asymptotic expressions, $\overline{C}_{kl}^{\pm} = \lim_{t \to
\infty} C_{kl}^{\pm} (t)$, into eq. (\ref{masterform}). This
replacement assumes that all internal dynamics of the chain
happens on much shorter time scales than the dynamics caused by
the interaction of the ancillas with the chain. Furthermore, it
does not treat the initial evolution for short times with full
accuracy since $\lim_{t \to 0} C_{kl}^{\pm} (t) = 0 (\not=
\overline{C}_{kl}^{\pm})$. The master equation thus obtained is
therefore valid in a regime where the couplings $J_a$ are weak
enough such that the time it takes for an excitation to travel
from $S$ to $R$ is completely determined by $J_a$, i.e. by the
time it takes to be transferred into and from the chain.

\begin{figure}
\psfrag{o}{\small{$\Omega_0$}}
\psfrag{f}{\small{$2 \, J_a^2 \, y_1$}}
\psfrag{6}{\hspace{-0.06cm}\raisebox{-0.2cm}{\small{$0.6$}}}
\psfrag{7}{\hspace{-0.06cm}\raisebox{-0.2cm}{\small{$0.7$}}}
\psfrag{8}{\hspace{-0.06cm}\raisebox{-0.2cm}{\small{$0.8$}}}
\psfrag{9}{\hspace{-0.06cm}\raisebox{-0.2cm}{\small{$0.9$}}}
\psfrag{0.0002}{\small{$ $}}
\psfrag{0.0004}{\hspace{-0.3cm}\small{$0.0004$}}
\psfrag{0.0006}{\small{$ $}}
\psfrag{0.0008}{\hspace{-0.3cm}\small{$0.0008$}}
\psfrag{0.001}{\small{$ $}}
\psfrag{0.0012}{\hspace{-0.3cm}\small{$0.0012$}}
\psfrag{0.0014}{\small{$ $}}
\includegraphics[width=7cm]{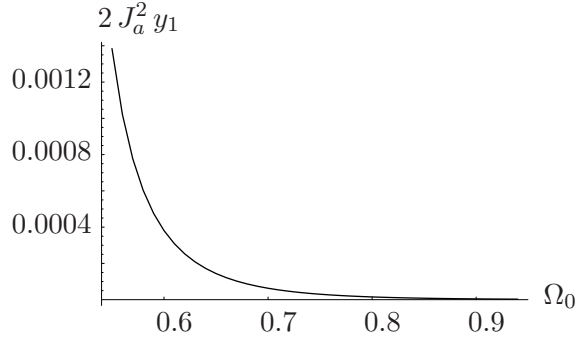}
\caption{\label{offresplot} Frequencies $2 J_a^2 y_1$ of the excitation's oscillation between
$S$ and $R$ for $\Omega = 1$, $J_a = 0.05$ and $\omega = 0.5$. The transfer speed increases
as $\Omega_0 - \omega \rightarrow 0$. Figure taken from \cite{HRP06}.}
\end{figure}

From eq. (\ref{masterform}) we find the following solution for the
expectation values of the occupation numbers of $S$ and $R$,
$n_S = \textrm{Tr}(a_S^{\dagger} a_S \sigma)$ and $n_R = \textrm{Tr}(a_R^{\dagger} a_R \sigma)$:
\begin{equation} \label{mastersol}
\lk.
\begin{array}{r}
n_S (t)\\
n_R (t)
\end{array}
\rk\} = \lk(  \frac{n_S(0) + n_R(0)}{2} \cosh(2 J_a^2 x_1 t) \pm
\frac{n_S(0) - n_R(0)}{2} \cos(2 J_a^2 y_1 t) \rk) \, e^{- 2 J_a^2
x_0 t}.
\end{equation}
Here, $x_0 = \lim_{t \to \infty} X_0$, $x_1 = \lim_{t \to \infty}
X_1$ and $y_1 = \lim_{t \to \infty} Y_1$. Note that $x_0 > 0$ and
$x_0 > |x_1|$. Inserting the correlation functions into eq.
(\ref{mastercoeff}) one sees that $x_0$ and $x_1$ are only non-zero if
$\omega \ge \omega_k$ for at least one mode $k$, that is if our
initial state is in resonance with states where both ancillas are
in their ground states and the chain is in one of its lowest-lying
excited states (c.f. figure \ref{energycon}). The dispersion
relation shows that this only happens for $\omega \ge \Omega_0$.
As in figures \ref{simulnonres} and \ref{simulres} or
\ref{simulnonres1} and \ref{simulres1}, we thus observe two
different scenarios:

If $\omega < \Omega_0$, and therefore $x_0 = x_1 = 0$,
the excitation that is initially in $S$ oscillates back and
forth between $S$ and $R$ at a frequency $2 J_a^2 y_1$, i.e.
\begin{equation} \label{mastersolnonres}
\lk.
\begin{array}{r}
n_S (t)\\
n_R (t)
\end{array}
\rk\} = \frac{n_S(0) + n_R(0)}{2} \pm \frac{n_S(0) - n_R(0)}{2} \cos(2 J_a^2 y_1 t) \, .
\end{equation}
Note in particular that the excitation is entirely transferred to $R$ at times
$t_n = n (\pi / J_a^2 y_1); \: n = 1, 2, \dots$ .
Figure \ref{offresplot} shows the frequencies $2 J_a^2 y_1$
of the excitation's oscillations between $S$ and $R$ for cases where $\omega < \Omega_0$,
for $\Omega = 1$, $J_a = 0.05$ and $\omega = 0.35$ as a function of $\Omega_0$.
As $\Omega_0 - \omega$ decreases, the transfer becomes faster and the oscillation frequency
increases.
%
%

Let us examine the dependence of the oscillation frequency $2
J_a^2 y_1$ on the distance $|m_S - m_R|$. With the definitions of
$y_1$ and $Y_1$ and eqs. (\ref{mastercoeff}), (\ref{correlfun})
and
    $\int_0^{\infty} d\tau \, e^{- i x \tau} =\pi \delta(0)- i \mathcal{P} \frac{1}{x} \,
    ,$
where $\mathcal{P}$ denotes the principal value, we find
\begin{equation}
2 J_a^2 y_1 = - \frac{J_a^2}{2 \pi \omega} \int_{- \pi}^{\pi} dk \,
\frac{\cos (|m_S - m_R| k)}{4 \Omega^2 \sin^2 (k / 2) + \Omega_0^2 - \omega^2} \, .
\end{equation}
Using the substitution $z = \ee^{\iu k}$, this integral can be
converted to an integral over the unit circle, which in turn can
be evaluated via the residue theorem \cite{AFcom}. The result is
\begin{equation}
2 J_a^2 y_1 = - \frac{J_a^2}{8 \omega \Omega^2} \,
\frac{z^{|m_S - m_R|} + z^{-|m_S - m_R|}}{\sqrt{\alpha^2 + \alpha}}
\end{equation}
where $z = 2 \sqrt{\alpha^2 + \alpha} - (2 \alpha + 1)$ and
$\alpha = (\Omega_0^2 - \omega^2) / 4 \Omega^2$.
The frequency $2 J_a^2 y_1$ and hence the transfer speed decrease exponentially with
increasing $|m_S - m_R|$.

\begin{figure}
\psfrag{t}{\raisebox{-0.4cm}{$t$}}
\psfrag{1}{\hspace{-0.4cm}\small{$1.0$}}
\psfrag{0.1}{\hspace{-0.1cm}\small{$0.0$}}
\psfrag{0.2}{\hspace{-0.1cm}\small{$0.2$}}
\psfrag{0.4}{\hspace{-0.1cm}\small{$0.4$}}
\psfrag{0.6}{\hspace{-0.1cm}\small{$0.6$}}
\psfrag{0.8}{\hspace{-0.1cm}\small{$0.8$}}
\psfrag{500}{\raisebox{-0.2cm}{\hspace{0.1cm}\small{$500$}}}
\psfrag{1000}{\raisebox{-0.2cm}{\hspace{0.1cm}\small{$1000$}}}
\psfrag{1500}{\raisebox{-0.2cm}{\hspace{0.1cm}\small{$1500$}}}
\psfrag{2000}{\raisebox{-0.2cm}{\hspace{0.1cm}\small{$2000$}}}
\includegraphics[width=7cm]{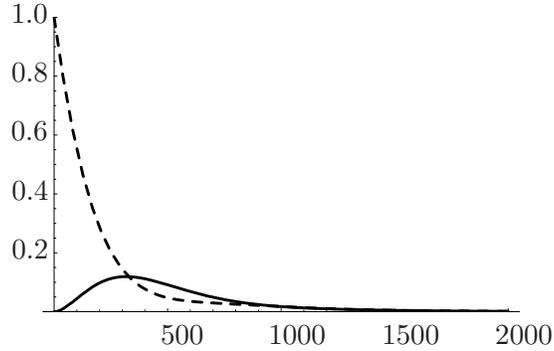}
\caption{\label{mastersolplot2} The solution (\ref{mastersol}) for a
harmonic chain and ancillas with $\Omega = 1$, $\omega = 0.5$, $J_a = 0.05$,
$|m_S - m_R| = 9$ and $\Omega_0 = 0.2$. Figure taken from \cite{HRP06}.}
\end{figure}

If, on the other hand, $\omega \ge \Omega_0$, the chain acts
similarly to a bath. Here, $x_0 \not= 0$, $x_1 \not= 0$, and both
ancillas relax into their ground state transferring their energy
into the chain. During this process, however, a fraction of the
energy initially located in $S$ appears momentarily in $R$
before it is finally damped into the chain.
The maximal excitation of the receiver throughout the entire evolution
depends on the distance $|m_S - m_R|$. For a given initial energy in $S$,
only a narrow range of the excitation spectrum is relevant for the dynamics.
Hence all relevant modes of the chain have very similar wavelength.
If the distance $|m_S - m_R|$ is a multiple of half that wavelength, a wave that
has high amplitude next to $S$ will have high amplitude next to $R$, too.
In that case, the maximal excitation probability in $R$ will be relatively large, although
less than $0.5$.
If on the other hand, a wave that has high amplitude next to $S$ has zero amplitude next to
$R$, the transfer vanishes completely.
Figure \ref{mastersolplot2} shows the solution (\ref{mastersol}) for a
harmonic chain and ancillas with $\Omega = 1$, $\omega = 0.5$, $J_a = 0.05$,
$|m_S - m_R| = 9$ and $\Omega_0 = 0.2$.
One might try to derive the same type of master equation for the
spin chain (\ref{hamchain}). In the next section, we explain why such an approach
is problematic when attempted with standard techniques. 

\subsection{The problem of even and odd excitation number subspaces for spin chains}

The standard procedure to diagonalize
say the transverse Ising chain is a Jordan Wigner transformation
followed by Fourier and Bogoliubov transformations \cite{LSM61}.
This procedure however only diagonalizes the subspace of odd or even
number of excitations at a time but not both.
Since the operator $\sigma_j^x$ maps between the odd to the even number
subspaces for any $j$, an exact expression for the time dependent correlation
functions $\bra 0 | \sigma_j^x (t) \sigma_l^x (s) | 0 \ket$, which would be required for
deriving a master equation, cannot be obtained
via the above procedure \cite{McC71}. 
If one uses only say the even excitation number subspace as an approximation,
the errors for the obtained eigenvalues scale as $1/N$, where $N$ is the length
of the chain. It is therefore often argued that only using one subspace is a good
approximation for long chains. 

\begin{figure}
\psfrag{t}{\hspace{0.2cm}\raisebox{-0.0cm}{$t$}}
\psfrag{r}{\raisebox{-0.4cm}{$ $}}
\psfrag{0.2}{\hspace{-0.1cm}\small{$0.2$}}
\psfrag{0.4}{\hspace{-0.1cm}\small{$0.4$}}
\psfrag{0.6}{\hspace{-0.1cm}\small{$0.6$}}
\psfrag{0.8}{\hspace{-0.1cm}\small{$0.8$}}
\psfrag{1}{\hspace{-0.4cm}\small{$1.0$}}
\psfrag{1.2}{\hspace{-0.1cm}\small{$1.2$}}
\psfrag{1.4}{\hspace{-0.1cm}\small{$1.4$}}
\psfrag{500}{\raisebox{-0.2cm}{\hspace{-0.1cm}\small{$500$}}}
\psfrag{1000}{\raisebox{-0.2cm}{\hspace{-0.1cm}\small{$1000$}}}
\psfrag{1500}{\raisebox{-0.2cm}{\hspace{-0.1cm}\small{$1500$}}}
\psfrag{2000}{\raisebox{-0.2cm}{\hspace{-0.1cm}\small{$2000$}}}
\includegraphics[width=7cm]{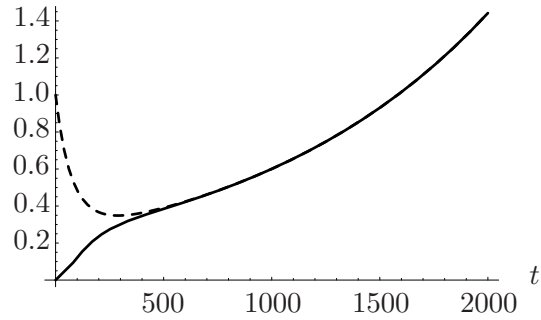}
\caption{\label{nonphys}
$P(\uparrow_S \downarrow_R)(t)$ (dashed line)
and $P(\downarrow_S \uparrow_R)(t)$ (solid line)
for $B = 1$, $J_x = 0.3$, $J_y = J_z = 0.0$, $B_a = 0.8$ and $J_a = 0.05$
as given by the solution of a master equation for the transverse Ising spin chain.
The solution is not physical since probabilities are not less or equal to 1.}
\end{figure}

In our problem, this reasoning cannot be applied: An approximation of the required
correlation functions, $\bra 0 | \sigma_j^x (t) \sigma_l^x (s) | 0 \ket$, by only
considering one subspace leads to unphysical results for the master equation.
An example is given in figure \ref{nonphys}.
We have therefore avoided a treatment of the spin chains with master equations. 

\section{Conclusions}

Quantum spin chains may be used as quantum channels. Here we have
studied the dependence of the transfer quality and speed on the
size of the spectral gap between the ground and the lowest excited
state of the considered system.

Two different scenarios are observed: Provided the spectral gap of
the chain is larger than the amount of energy initially located in
the sender, close to perfect transfer happens on rather long time
scales. The transfer becomes even more perfect but slower if the
coupling strength between ancillas and chain or the energy
initially in the sender is lowered further.

If the energy in the sender is initially larger than the spectral
gap, the behavior changes significantly. Here, the excitations get
lost into the chain implying a bad transfer quality. This second
scenario takes place on much shorter time scales than the first
one.

These results may be used for two main applications: For quantum
information purposes, they show, that by using a system with a
large enough spectral gap one can build good quantum channels.
These channels are furthermore considerably robust against
decoherence because the chain remains to a high precision in its
ground state which is not significantly affected by an environment
at zero temperature.

In condensed matter physics, on the other hand, the transfer speed and quality can
be used to infer the size of the spectral gap and hence the location of
quantum critical points. This method is admittedly inferior to neutron scattering
in magnetic composites, but it may by advantageous in optical lattices where neutron scattering
is not that successfull.

\begin{acknowledgments}
The authors would like to thank Sougato Bose and Daniel Burgarth
for discussions at early stages of this project as well as
comments by Andrew Fisher. This work is part of the QIP-IRC
supported by EPSRC (GR/S82176/0), the Integrated Project Qubit
Applications (QAP) supported by the IST directorate as Contract
Number 015848' and was supported also by the Alexander von
Humboldt Foundation, Hewlett-Packard and the Royal Society.
\end{acknowledgments}


\begin{thebibliography}{99}
\bibitem{AFcom}
We like to thank Andrew Fisher for bringing this point to our attention.
\end{thebibliography}
\end{document}